\documentclass[manuscript]{aastex}

\slugcomment{Not to appear in Nonlearned J., 45.}

\shorttitle{Coronal Jet above Light Bridge} \shortauthors{S. Liu et
al.}

\begin{document}


\title{A Coronal Jet Ejects from Sunspot Light Bridge}

\author{S. Liu\altaffilmark{1}}

\affil{National Astronomical Observatory, \\Chinese Academy of
Sciences,
        Beijing, China}

\email{lius@nao.cas.cn}

\altaffiltext{1}{key Laboratory of Solar Activity}


\begin{abstract}
Chromospheric brighten and H$\alpha$ surge are the evident and
common phenomena along sunspot light bridge. In this paper, a
coronal jet ejects from sunspot light bridge is presented. Using the
data from the Solar Dynamics Observatory (SDO) and Hinode
satellites, it is confirmed that the jet has the root near light
bridge, this suggests that the jet may be a result of reconnection
between main sunspot and light bridge. Due to the processing of jet
ejects, the intensity and width of light bridge have some changes at
some extent. This also suggests that jet is related to the
interaction between light bridge and umbra, possibly magnetic
reconnection or heat plasma trapped in light bridge escaping and
moving along field line.
\end{abstract}


\section{Introduction}

Light bridge (LB) is bright structures crossing the umbra during the
evolution of sunspots. LB is associated to the breakup of sunspots
in the decay or the assembly of sunspots in complex active regions
\citep{bra64, vas73, gar87}. \citet{mul79} classified LB as
"photospheric,""penumbral," and "umbral" LB according its intensity
and fine structure. Another classification is as follows
\citep{sob93, sob94}: strong LB, which separates umbral core and is
further distinguished as photospheric or penumbral, and faint LB,
which is faint narrow lane with in the umbra and most likely
consists of umbral dots.

At present, the formation and magnetic properties of LB are not
understood completely. A common mechanism to explain the formation
of LB is that field-free convection penetrates umbra from
sub-photosphere and forms a cusp-like magnetic field \citep{spr06}.
\citet{kat07b} revealed the formation of a LB due to the intrusion
of umbral dots basing on data obtained from H$inode$ satellite.
Magnetic field in LB is revealed weaker and more inclined than that
of surround umbra \citep{rue95, lek97, jur06}. \textbf{Based} on
H$inode$ observation of the magnetic field in a LB accompanied by
long-lasting chromospheric plasm ejections, \citet{shi09} suggest
that current-carrying highly twisted magnetic flux tubes are trapped
below a cusp-shape magnetic structure along the LB. In addition, by
a detail analysis of the Stokes spectra \citep{jur06}, it is found
that the field strengths and inclinations increase and decrease with
height, which suggest a canopy-like structure above the LB.

It is undoubted that the plasm contained in LB have a temperature
higher than that in surrounding umbra because of the brightness of
LB. Observations indicate that there are remarkable plasma ejections
or H$\alpha$ surge activities in chromosphere along LB \citep{roy73,
asa01, bha07, shi09}. Additionally, over the site of LB the bright
enhancement in 1600 \AA ~images and heating of coronal loops in 171
\AA ~images from Transition Region and Coronal Explorer
(\textbf{TRACE}) was founded recently \citep{ber03, kat07a}, which
observations suggest that LB is a steady heat source in the
chromosphere. \citet{lou08} using G-band and Ca II images obtained
from H$inode$ studied dynamics and brightness enhancements of LB ,
and pointed to the possibility that LB could be the sites for
heating the overlying chromosphere, but can not rule out the
likelihood of coronal phenomena. In this paper, a coronal jet
originates from the site of LB is presented, which means that the
interaction between LB and umbra can also create coronal dynamic
activities.

The paper is organized as follows: firstly, the description of
observations and data used will be introduced in
Section~\ref{S-Obser and Data}; secondly, the results are shown in
Section~\ref{S-Results}; at last, the short discussions and
conclusions will be given in section~\ref{S-Conl}.

\section{Observations and Data Reduction}
\label{S-Obser and Data}

Jet studied here is near disk center (heliographic coordinates
~S17W23), it occurred during about 20:00-21:00 UT on 29 Mar 2011.
The observatory data, which used to study this jet, were obtained by
Atmospheric Imaging Assembly (AIA) \citep{tit06, boe10} and
Helioseimic and Magnetic Imager (HMI) \citep{sch10} on board the SDO
and by Solar Optical Telescope (SOT) on board H$inode$ \citep{kos07,
tsu08}. AIA takes full-disk image in 10 wavelengths with a pixel
size of 0.59 arcsec. In this paper, AIA 171~\AA~ and 1700~\AA~ data
at Level 1 with 45 s cadence is used. HMI includes full-disk
magnetograms, continuum intensities, dopplergrams with spatial
resolution of 0.5 arcsec. In this paper, light-of -sight (LOS)
magnetograms and continuum intensity are used for this analysis.
G-band and Ca II (with spatial resolution of 0.1 arcsec) and LOS
magnetogram (with spatial and temporal resolution 0.16 arcsec and 15
min, respectively) observed by SOT/H$inode$ are used in this work.
The data processing in the work are all based on standard solar
software (SSW {\it e.g}, fg\_prep.pro, aia\_prep.pro). For example,
dark subtraction, flat fielding, the correction of bad pixels and
cosmic-ray removal were done for filtergram images obtained by SOT.
\section{Results}
\label{S-Results}

\begin{figure}

   \centerline{\includegraphics[width=1.\textwidth,clip=]{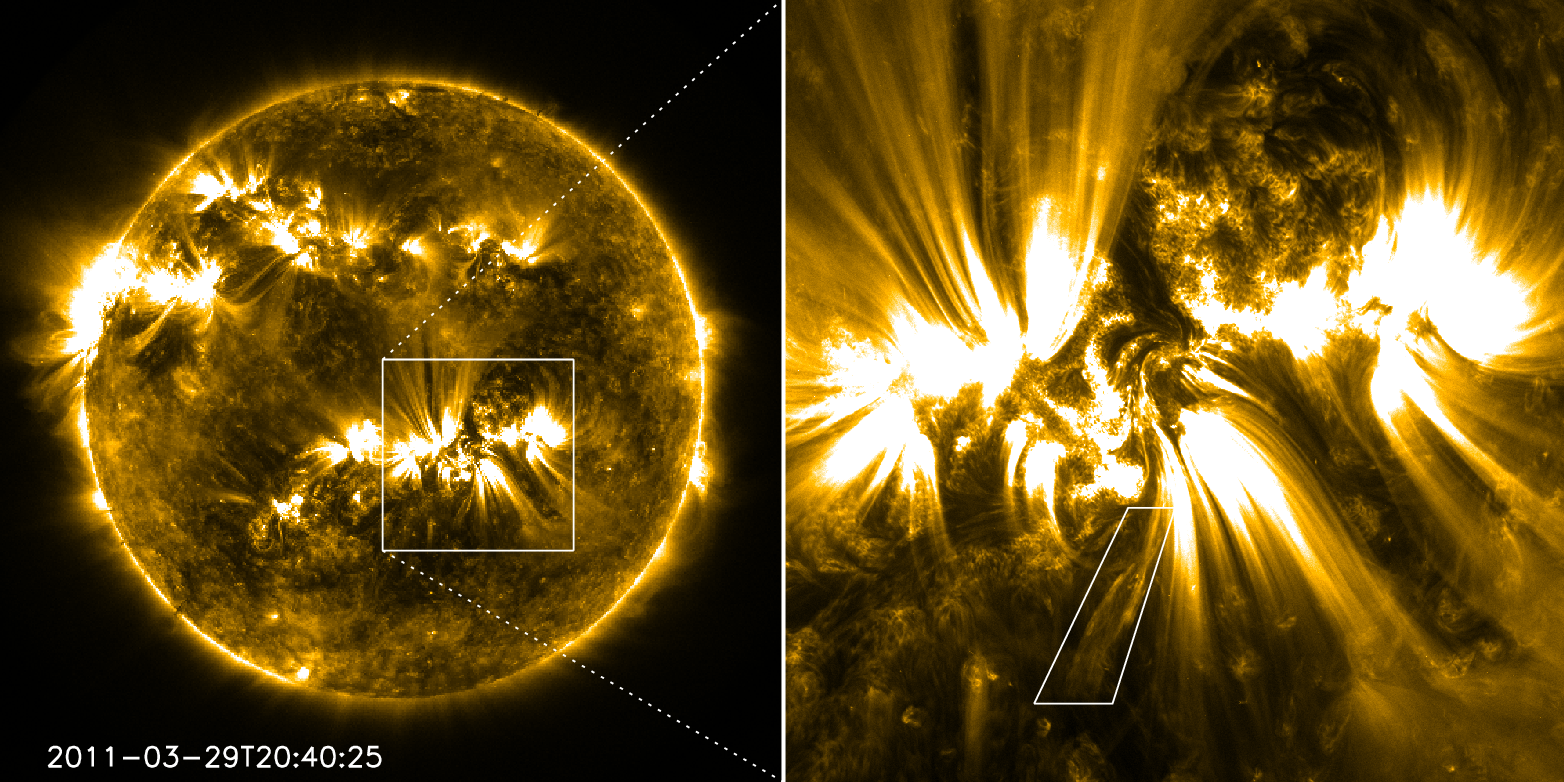}}

   \caption{The left image is full-disk AIA 171 image at 20:40;25 UT on 03 29 2011.
   The right one is sub-region drawn in the left image, where coronal jet
   is highlighted by a rectangle drawn in right image.
} \label{Fig1}
\end{figure}

Fig~\ref{Fig1} shows the location of this jet, it occurs during
about 20:00-21:00 UT on 29 Mar 2011. The left image is full-disk AIA
171~\AA~ image at 20:40 UT, and the region where jet occurs is drawn
by a white rectangle, the right image is amplified sub-region of
left image, where the jet (highlight by white region) can be seen
obviously. It is found that the jet has already separated in two
parts along its direction of propagation as time going on, the
former part ejects toward space and the latter one falls back along
magnetic field lines where it created.

\begin{figure}
   \centerline{\includegraphics[width=1.\textwidth,clip=]{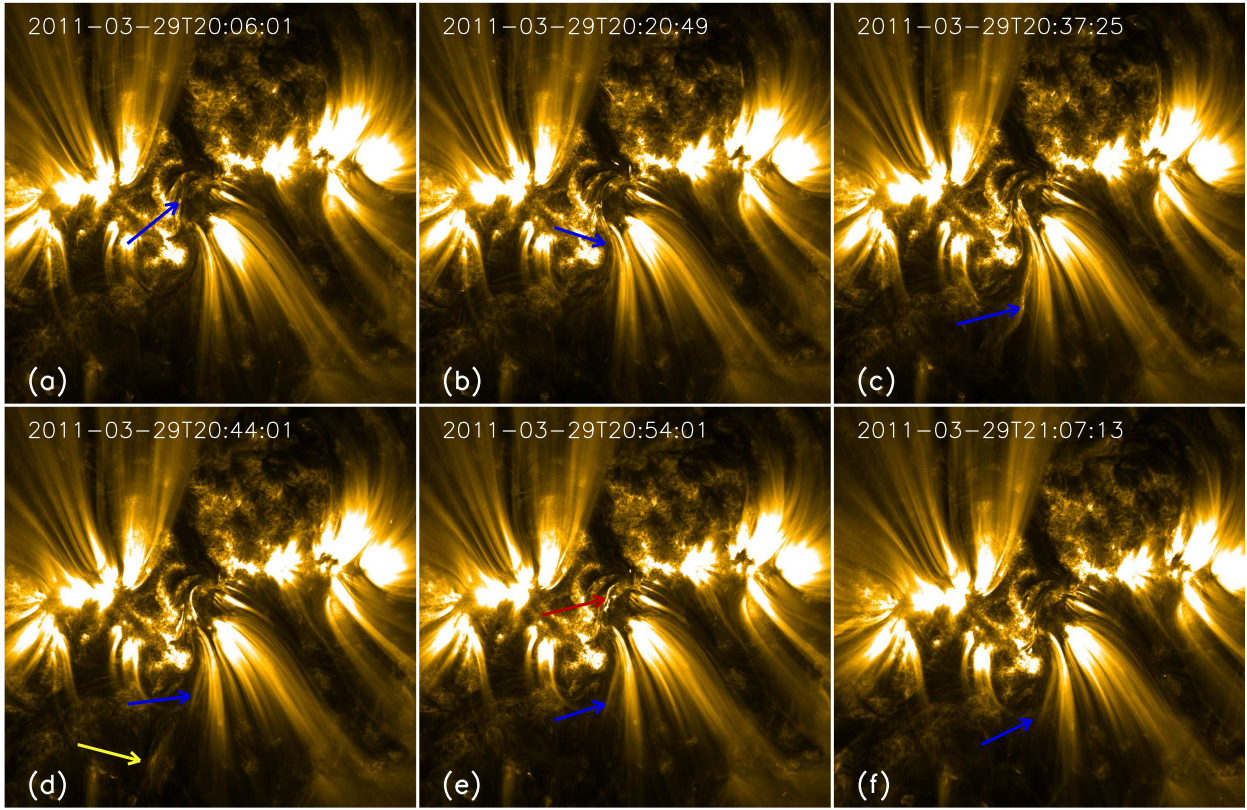}}

   \caption{Time series images of AIA 171 channel which show the process of jet. It includes the start (a, b),
maximum (c, d) and decay (e, f) phases. It can be seen in image (d)
that the jet have already separated two parts along its direction of
propagation.} \label{Fig2}
\end{figure}

\begin{figure}
   \centerline{\includegraphics[width=1.\textwidth,clip=]{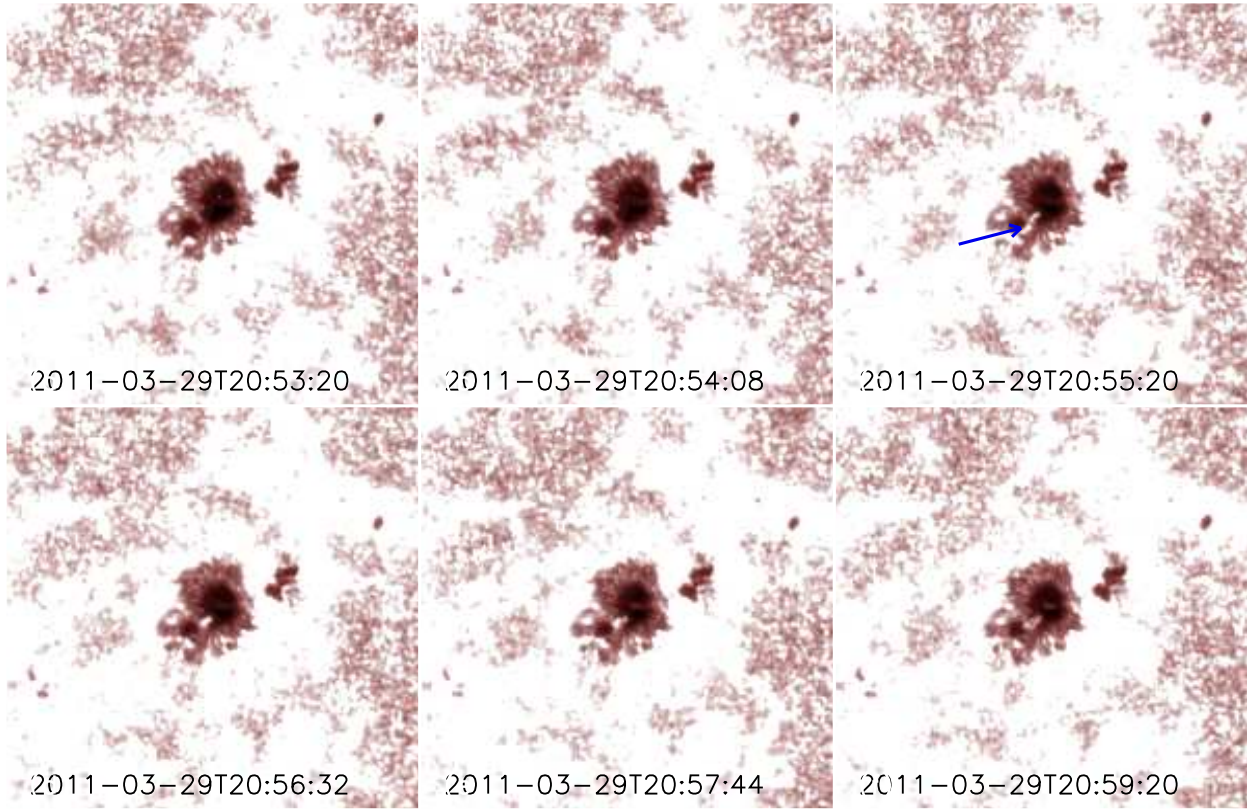}}

   \caption{The evolution of 1700 \AA~ images from 20:53
to 20:59 UT, the blue arrow in panel (e) should be caused by the
back-flow of jet. } \label{Fig21}
\end{figure}

Fig~\ref{Fig2} shows the process of coronal jet from 20:00 to 21:07
UT, with peak at 20:45 UT using AIA 171~\AA~ images. The field of
view is 162 arcsec for all images. The blue arrow in each frame
points to the evolution of this jet, includes the start (a, b),
maximum (c, d) and decay (e, f) phases. In image (d) the jet has
already separated two parts along its direction of propagation, the
former part(yellow arrow) ejects toward space away from the Sun, the
latter one (blue arrow) becomes a back-flow along magnetic field
lines and gives rise to the intensity enhancement of the site where
jet originated, which can be seen in image (e) indicates by a red
arrow. The velocity of back-flow is about 198 km/s, which is
calculated from these AIA 171~\AA~ images, however the effect of
project are not considered. To see the intensity enhancement
resulted from the back-flow, Fig~\ref{Fig21} shows the evolution of
1700 \AA~ images from 20:53 to 20:59 UT. The intensity enhancement
indicated by a blue arrow in panel (e) should be caused by the
back-flow corresponding the latter part of jet.

\begin{figure}
   \centerline{\includegraphics[width=1.\textwidth,clip=]{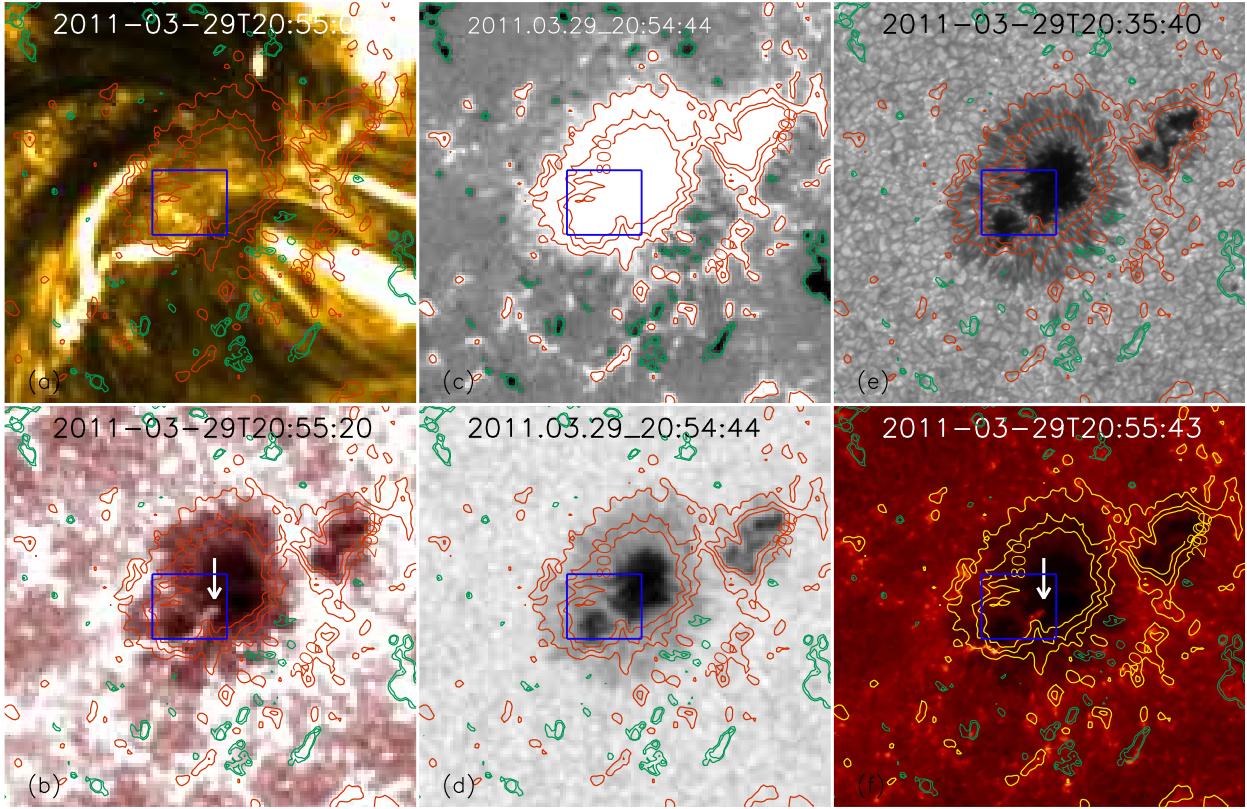}}

   \caption{The 171~\AA~ (a) and 1700 \AA~ (b) AIA images, LOS magnetogram (c) and continuum intensity
(d) HMI images and G-band (e) and Ca II (f) H$inode$ images. The
contours drawn on each image are light-of-sight magnetic field,
where the levels of magnetic field contours are -80 to -50 and
200-800 G. } \label{Fig3}
\end{figure}

\begin{figure}
   \centerline{\includegraphics[width=1.\textwidth,clip=]{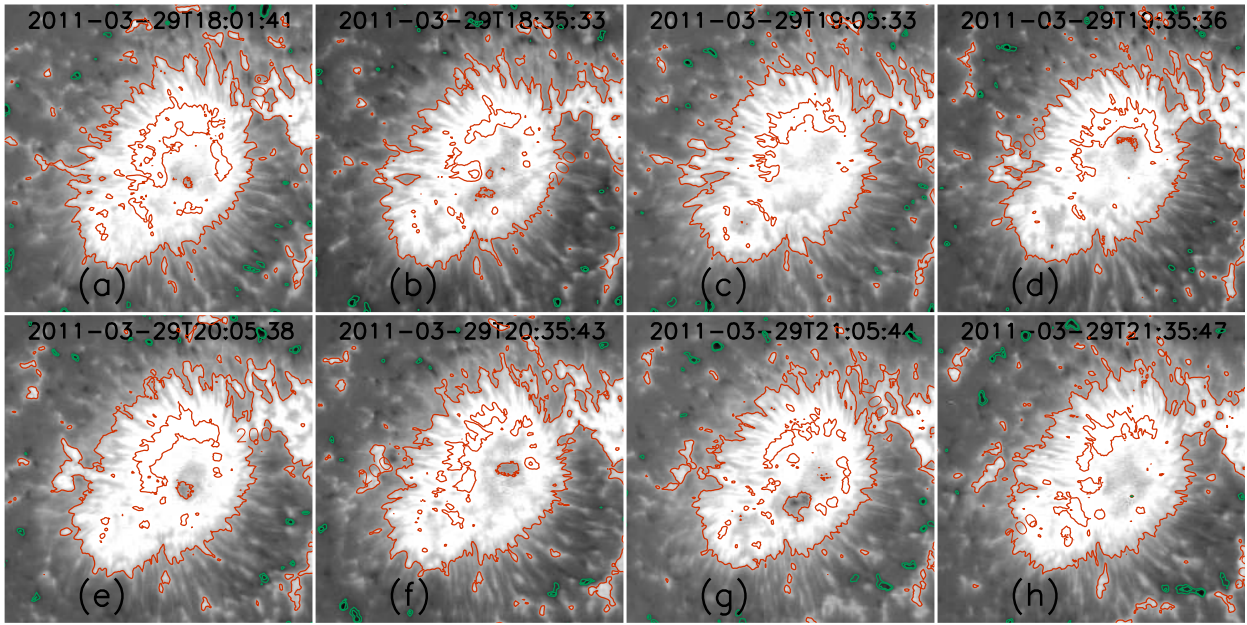}}

   \caption{The evolution of LOS magnetic field observed by SOT/H$inoed$ during 18:01-21:35 UT.
The contours levels are from -80 to -50 (green) and 200-800 (red) G,
and the field of view is 40 $\times$ 40 arcsec$^2$. From these
images it is found that comparing the magnetic flux of main sunspot
there is no opposite magnetic flux near LB. } \label{Fig4}
\end{figure}

In order to know the circumstances of lower atmosphere corresponding
to the site where jet created, Fig~\ref{Fig3} gives the AIA 171~\AA~
(a) and 1700 \AA~ (b) images, HMI LOS magnetogram (c) and continuum
intensity (d) images and G-band (e) and Ca II (f) SOT/H$inode$
images together. These images are aligned by heliospheric coordinate
and combined by correlation of feature points (cross-correlation
function IDL). The contour lines plotted on each image are LOS
magnetic field, where the levels of magnetic field contours are -80
to -50 and 200-800 G, and red/green contours are positive/negative
magnetic flux, while in image (f) the yellow contour are positive
one, the blue rectangle indicates the site of jet originated. From
the continuum intensity, G-band and 1700 \AA~ image, it can be seen
clearly that the jet has its root near LB, which can be seen from
the intensity enhancement of image (b) and (f) indicated by a white
arrow, this intensity enhancement is due to mass falling back along
magnetic field lines. Additionally, from image (c) it can be found
that there is no evident negative magnetic flux, which opposite the
main positive magnetic flux nearby the root of jet. To see the fine
structure of magnetic field on the photospheric associated with jet
eruption more clearly, the high spatial resolution (0.16 arcsec) LOS
magnetic field observed by SOT/H$inode$ is shown in Fig~\ref{Fig4},
here the field of view is 40 $\times$ 40 arcsec$^{2}$. Where the
levels of magnetic field contours are -80 to -50 and 200-800 G, and
red/green contours are positive/negative magnetic flux, which is
consistent with those of HMI LOS magnetic field in Fig~\ref{Fig4}.
It is found that there is no evident negative magnetic flux near the
site where jet eruption from the high spatial resolution images of
LOS magnetic field.

\begin{figure}
   \centerline{\includegraphics[width=1.\textwidth,clip=]{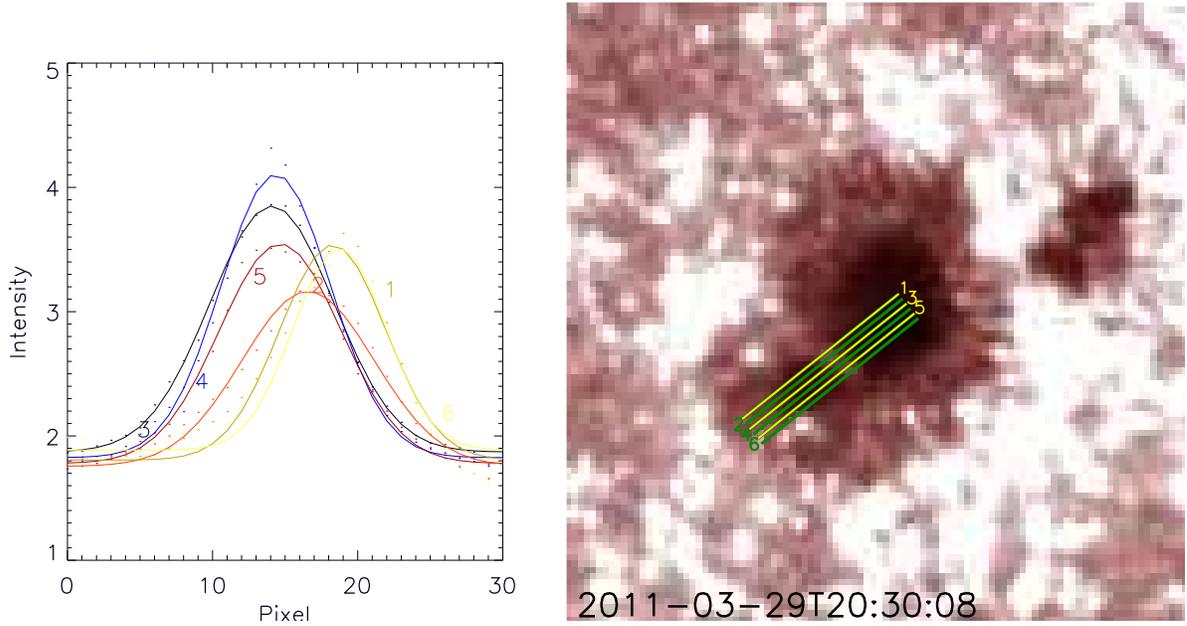}}

   \caption{The right frame is example of 1700\AA image, shows six selected slits
   along LB. The left frame plots the intensity of slits, and the
   profile of intensity obtained from Gaussian fit. The dot line is
profile of intensity and solid line is Gaussian fit, the numbers of
slit are corresponding those of the profile of intensity labeled }
\label{Fig5}
\end{figure}

In order to find the change in LB during jet ejected, the evolution
of maximal intensity and width of LB are studied using 1700~\AA~
images. Six slits are selected along LB (the right image of
Fig~\ref{Fig5}, which is an example observed at time 20:30:08 UT).
The profile of intensity of each slit is fitted using Gaussian
function, then the half width and the height of Gaussian function is
as the width and maximal intensity of LB, respectively (in left
image of Fig~\ref{Fig5}, the dot line is profile of intensity and
the solid line is Gaussian fit to the intensity, the numbers of slit
are corresponding those of the profile of intensity labeled using
different color). It can be found that the results of Gaussian fit
are reasonable and acceptable, because there is no evident
deviations between the true profiles and the fitted profile.

\begin{figure}
   \centerline{\includegraphics[width=1.\textwidth,clip=]{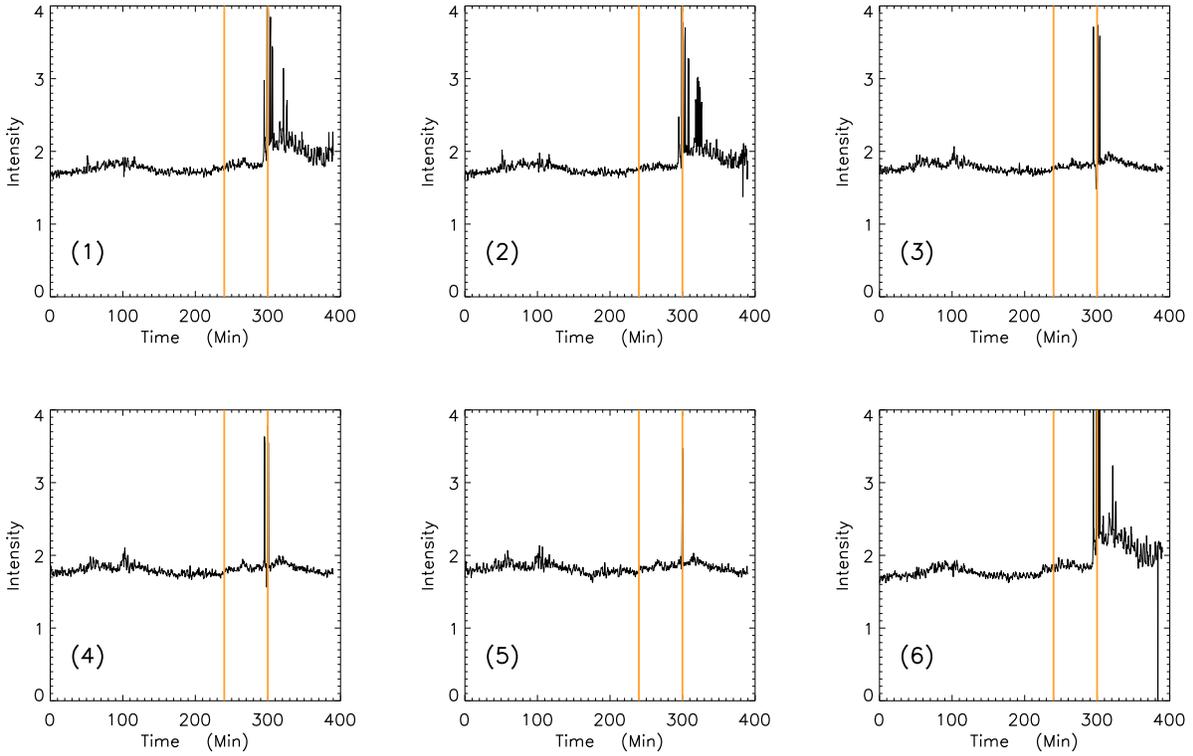}}

   \caption{The evolution of maximal intensity along LB obtained
    from Gaussian fit of six slits in AIA 1700~\AA images. The time
    interval between two yellow vertical lines is 20:00-21:00 UT.
} \label{Fig6}
\end{figure}

\begin{figure}
   \centerline{\includegraphics[width=1.\textwidth,clip=]{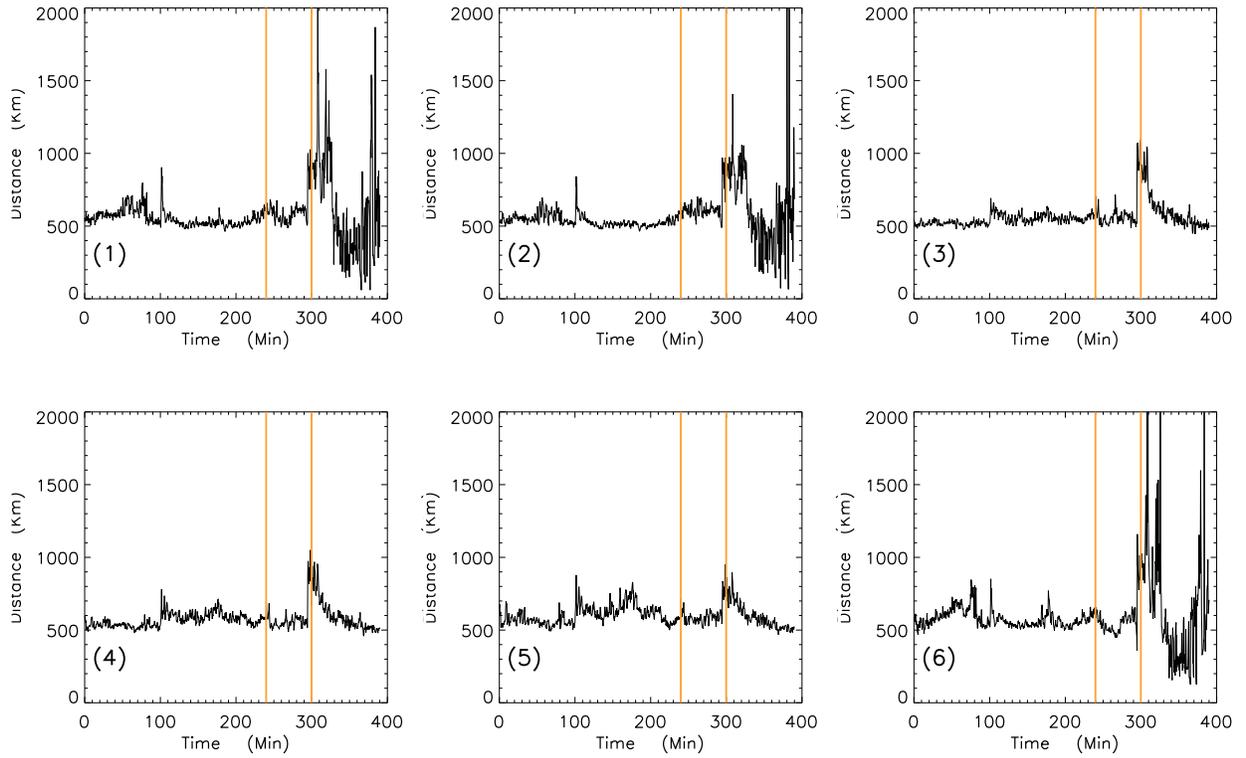}}

   \caption{The evolution of width of LB obtained
   from Gaussian fit of six slits in AIA 1700~\AA images. The time
   interval between two yellow vertical lines is 20:00-21:00 UT.
} \label{Fig7}
\end{figure}

Using the above method to calculate a serial of 1700~\AA~ images,
the evolution of maximal intensity and width of LB are plotted in
Fig~\ref{Fig6} and~\ref{Fig7}, respectively, during 16:00-22:30 UT.
The time interval between two yellow vertical lines is 20:00-21:00
UT. From Fig~\ref{Fig6}, it can be found that before jet creates
there is no evident changes of maximal intensity; by the time of jet
onset the maximal intensity begin to increase slightly, but this
increase is not evident; during the jet ejects the evolutions of
maximal intensity are different among six slits, for slit of (1) (2)
and (3) it increase slightly but not evidently, for slit of (4), (5)
and (6) it do not change almost, however on the whole they have a
trend of increase for maximal intensity during jet ejects; when jet
finishes the maximal intensity reach to maximum, which is caused by
the back-flow of jet, after that time it begin to decrease. From
Fig~\ref{Fig7}, it can be found that before and during jet ejects
there is no evident rule can be obtained, however after jet finishes
there is a jump of the width of LB, which is also caused by the
back-flow of jet, and then narrows down to normal width as before
quickly. The above results confirm that this jet has its root near
LB, thus, the interaction between LB and main sunspot may be a
direct reason for the creation of this jet. The width of LB
($\approx$ 500 Km) plotted in Fig~\ref{Fig7} is consistency the
previous result \citep{lou08}.

\begin{figure}
   \centerline{\includegraphics[width=1.\textwidth,clip=]{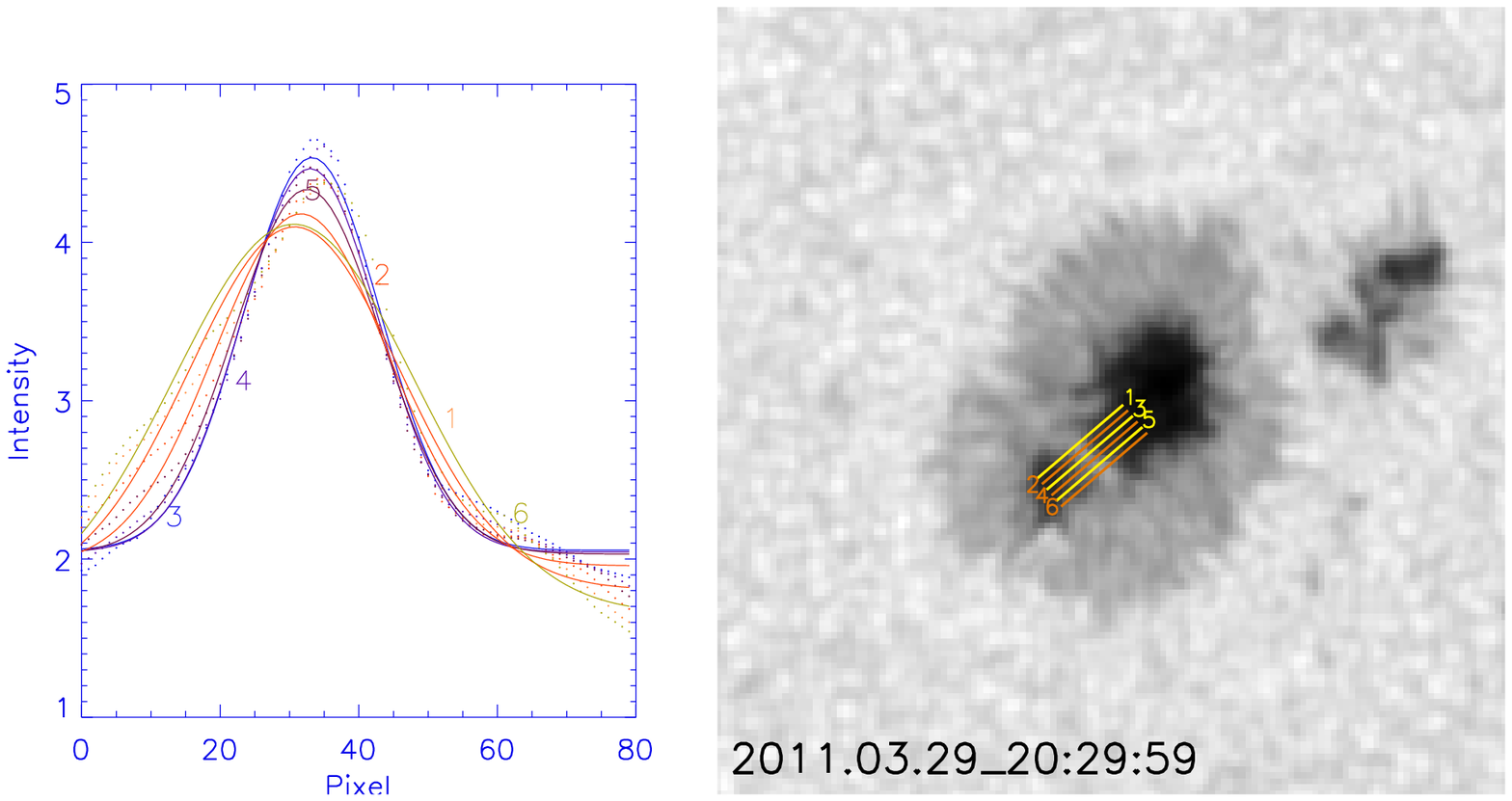}}

   \caption{Same as Fig~\ref{Fig4}, but for HMI continuum intensity.
} \label{Fig8}
\end{figure}

\begin{figure}
   \centerline{\includegraphics[width=1.\textwidth,clip=]{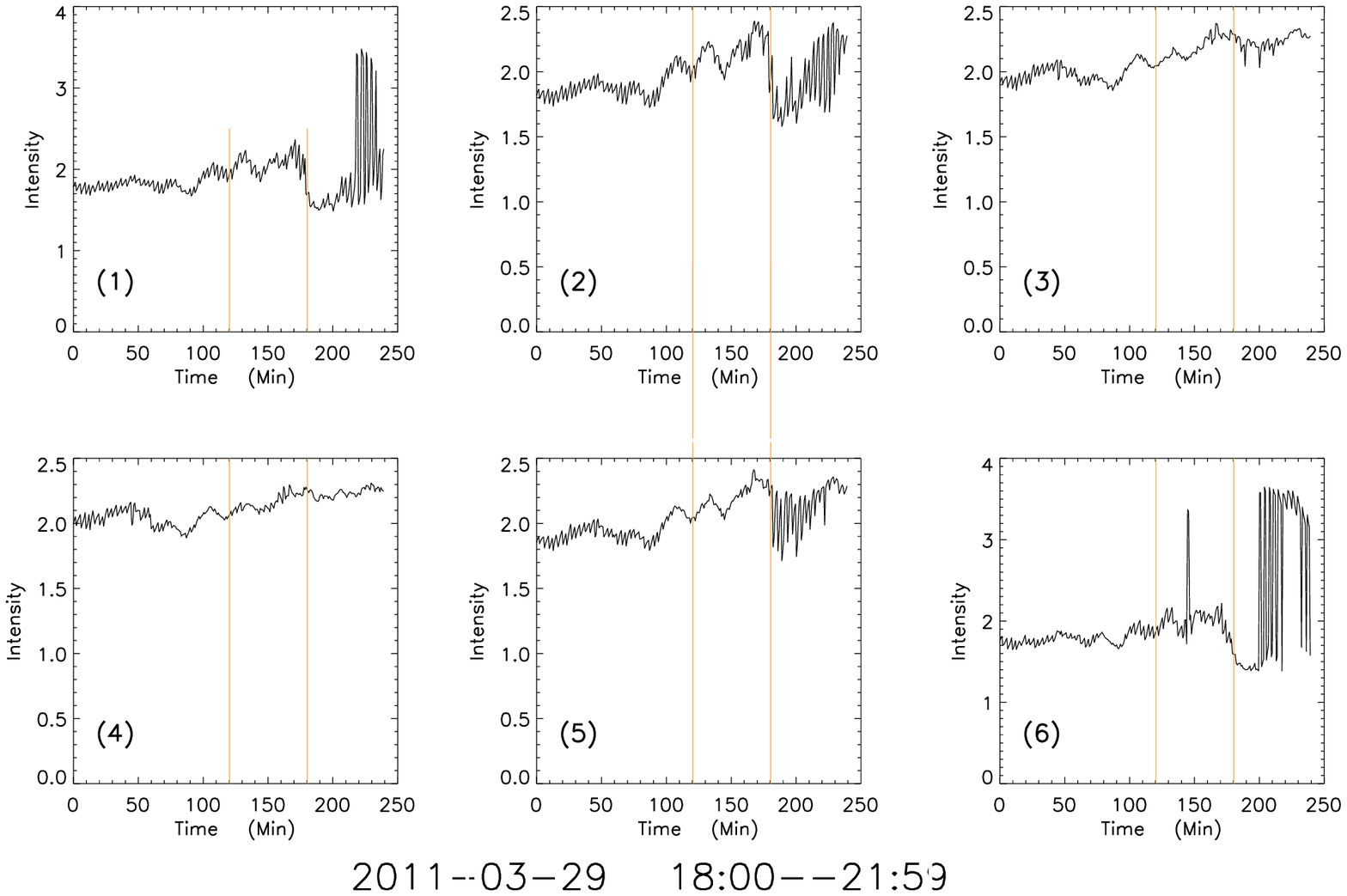}}

   \caption{Same as Fig~\ref{Fig5}, but for HMI continuum intensity.
} \label{Fig9}
\end{figure}

\begin{figure}
   \centerline{\includegraphics[width=1.\textwidth,clip=]{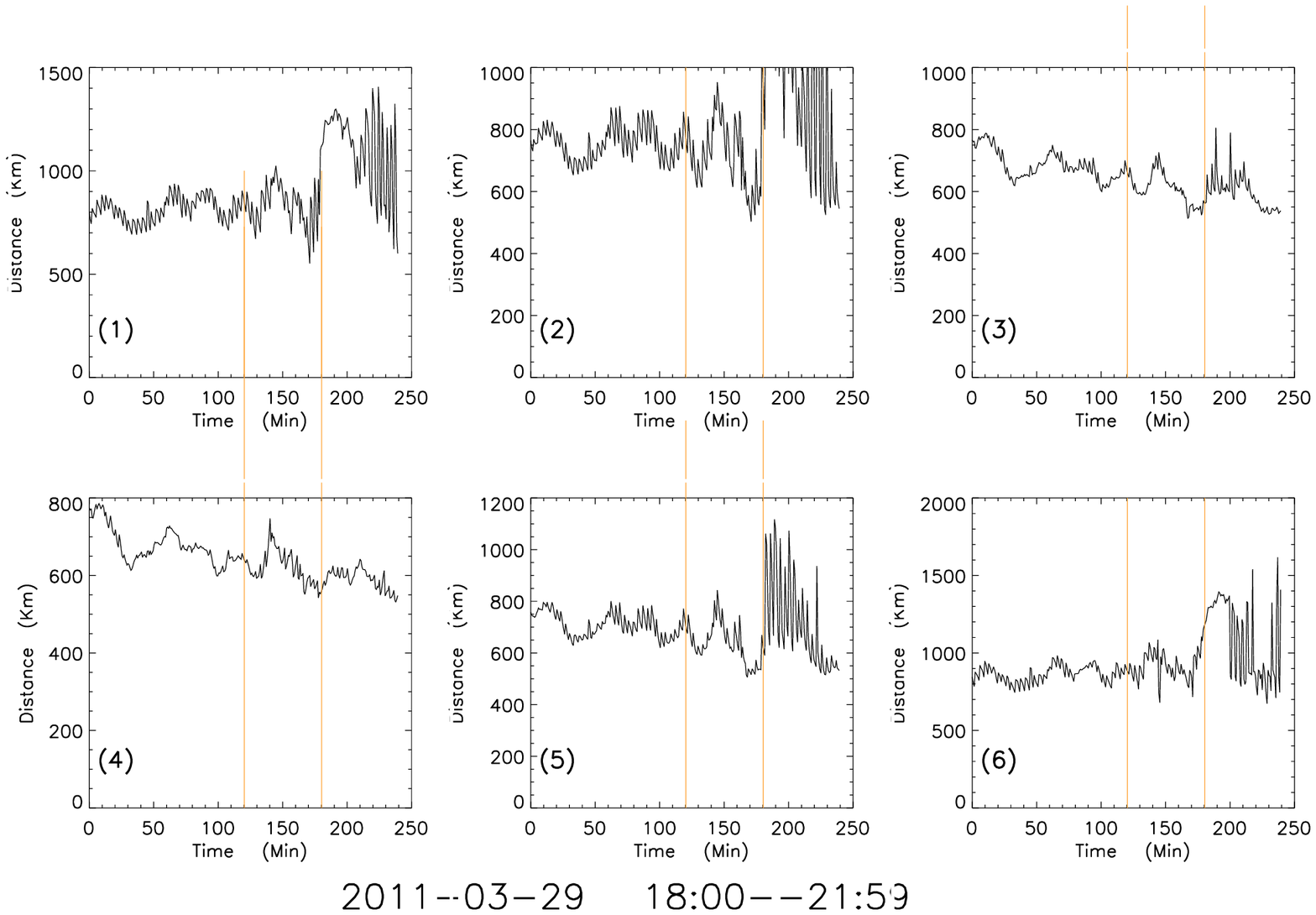}}

   \caption{Same as Fig~\ref{Fig6}, but for HMI continuum intensity.
} \label{Fig10}
\end{figure}

The HMI continuum intensity images are also studied by the same
method as above to know the properties of LB on photosphere during
jet erupts. Similar to Fig~\ref{Fig5}, six slits in continuum
intensity image are selected and shown in the right frame of
Fig~\ref{Fig8}, where the observation obtained at the time 20:29:59
UT is used as an example. After Gaussian fit for a serial of
continuum intensity images, the evolution of maximal intensity and
width of LB on the photosphere are plotted in Fig~\ref{Fig9} and
~\ref{Fig10}, respectively. On the whole, the maximal intensities
increase slightly before and during jet ejects, then they displays a
decrease after jet finishes. On the photosphere the change trend of
the width of LB can not be rule out before and during jet eject, and
different slits shows different change. However there is a common
fact that the width of LB is broaden at some extent after jet
finishes. It is also found that the width of LB on the photosphere
($\approx$ 800 Km) is broader than that of on 1700\AA~ image
($\approx$ 500 Km).

\section{Discussions and Conclusions }
\label{S-Conl}

Using multi-spectral images observed by H$inode$ and SDO newly
launched satellites, the evolution of a LB accompanying coronal jet
eruption is studied.It is reveal that this coronal jet (ejected
during 20:00 to 21:07 UT on Mar 29 2011, with peak at 20:45 UT) is
related to LB. It suggest that interaction between LB and main umbra
not only have low atmospheric response (previous studies include
H$\alpha$ surge, coronal loops enhancement) but can also has more
dynamic high atmospheric or coronal activities.

The evolutions of LB during jet eruption are studied basing on the
1700\AA~ images and photospheric continuum intensity images in this
paper. On the whole the intensity and width of LB show no evident
change before and during jet eject, however there are evident change
after jet finishes, which means that LB can also has dynamic coronal
response not only chromospheric response (chromospheric brighten and
H$\alpha$ surge). It is also find that the width of LB displayed on
the photosphere is broader than that displayed on 1700~\AA~ image.
The evolutions of intensity and width of LB at different atmosphere
can only give us a change trend during jet ejects. It is noted that
LB companying coronal jet is seldom for study so far. Thus, it is
expected that more available scientific data for similar analysis in
not far future.

Normally jet is regard as the phenomenon of magnetic reconnection.
The exist of opposite magnetic flux is required for reconnection
models, but from light-of-sight magnetic field it is found that
there is no evident opposite magnetic flux exist in the region where
jet originated for this event. Thus, there should be the magnetic
components in LB that can provide the essential condition for
magnetic reconnection. Accurately, the basic condition for magnetic
reconnection is the opposite magnetic flux and anomalous resistance.
Generally, the opposite magnetic flux in fact is two magnetic
topologies and the anomalous resistance is usually caused by some
instabilities. Here, the main sunspot and the LB can be considered
two magnetic systems and the instability maybe more easy exist at
the boundary between main sunspot and LB. For this jet event, the
opposite magnetic flux maybe below the resolution of magnetic field
observed. Or likes \citet{shi09}, highly twist magnetic field is
trapped in LB that can provide condition for magnetic reconnection.
Also it should be noted that magnetic reconnection do not always
require the opposite magnetic flux, such as component reconnection
(there are some angle differences between the two magnetic
components). However there may be more deep physics mechanism for
magnetic reconnection, which can be studied in future, when we get
high spatial and temporal resolution vector magnetic field that can
show more fine magnetic structure of LB.

Plasma and magnetic field fill the whole Sun, and there is a
phenomenon of magnetic freezing in the region of strong magnetic
field (namely, the region of sunspot). Hence, the plasma should flow
along magnetic field line. Sunspot and LB should be considered as
two magnetic systems. Previous studies \citep{rue95, lek97, jur06}
show that magnetic field in LB is weaker and more inclined than that
of surround umbra. For a unipolar sunspot studied here, the magnetic
field line should have radial shape, and the stronger magnetic field
the more vertical. LB is another magnetic system, the distribution
of magnetic field line should be more inclined. Thus, the plasma
flow the field line of individual magnetic system of sunspot and LB.
The fine topology of magnetic field of LB is unknown for us due to
the restrictions of observatory. Maybe the field lines along the
axis of LB or field lines surround the axis of LB. Hence the plasma
maybe flow along the axis of LB or they surround the axis of LB.
Either field lines along LB or they surround LB, they should be
considered as another magnetic system comparing to that of main
sunspot, which different from the main sunspot. Thus, there can
exist the basic condition for magnetic reconnection at the boundary
between sunspot and LB, and jet unavoidable become the results of
magnetic reconnection.
 \acknowledgments

{\it Hinode} is a Japanese mission developed and launched by
ISAS/JAXA, collaborating with NAOJ as a domestic partner, NASA and
STFC (UK) as international partners. Scientific operation of the
{\it Hinode} mission is conducted by the {\it Hinode} science team
organized at ISAS/JAXA. This team mainly consists of scientists from
institutes in the partner countries. Support for the post-launch
operation is provided by JAXA and NAOJ (Japan), STFC (U.K.), NASA,
ESA, and NSC (Norway). This work was partly supported by the
 National Natural Science Foundation of China (Grant Nos. 10611120338,
 10673016, 10733020, 10778723, 11003025,11103037 and 10878016),
 National Basic  Research Program of China (Grant No. 2011CB8114001) and Important
 Directional Projects of Chinese Academy of Sciences (Grant No. KLCX2-YW-T04).

\email{aastex-help@aas.org}.

\end{document}